# Exploring the transport properties of equatorially low-coordinated erbium single ion magnets

Silvia Giménez-Santamarina[a], Salvador Cardona-Serra [a*], José J. Baldoví[b*]

[a]*Instituto de Ciencia Molecular, Universitat de València, C/Catedrático José Beltrán 2, E-46980 Paterna, Spain.*

[b]*Max Planck Institute for the Structure and Dynamics of Matter, Luruper Chaussee 149, 22761 Hamburg, Germany.*

*Corresponding authors.
E-mail addresses: salvador.cardona@uv.es (S. Cardona-Serra), jose.baldovi@mpsd.mpg.de (J.J. Baldoví)

**ABSTRACT**

Single-molecule spin transport represents the lower limit of miniaturization of spintronic devices. These experiments, although extremely challenging, are key to understand the magneto-electronic properties of a molecule in a junction. In this context, theoretical screening of new magnetic molecules provides invaluable knowledge before carrying out sophisticated experiments. Herein, we investigate the transport properties of three equatorially low-coordinated erbium single ion magnets with $C_{3v}$ symmetry: Er[N(SiMe$_3$)$_2$]$_3$ (**1**), Er(btmsm)$_3$ (**2**) and Er(dbpc)$_3$ (**3**), where btmsm = bis(trimethylsilyl)methyl and dbpc = 2,6-di-*tert*-butyl-*p*-cresolate. Our ligand field analysis, based on previous spectroscopic data, confirms a ground state mainly characterized by $M_J$ =±15/2 in all three of them. The relaxation of their molecular structures when placed between two Au (111) electrodes leads to an even more symmetric ~$D_{3h}$ environment, which ensures that these molecules would retain their single-molecule magnet behavior in the device setup. Hence, we simulate spin dependent transport using the DFT optimized structures on the basis of the non-equilibrium Green's function formalism, which, in **1** and **2**, suggests a remarkable molecular spin filtering under the effect of an external magnetic field.

## 1. Introduction

Spintronics (spin-based electronics) has wrought a huge revolution in electronic data processing since the discoveries of giant[1] and tunnel magnetoresistance[2]. The main goal of this scientific research field is the active manipulation of the spin degree of freedom of carriers, e.g. electrons.[3] Spintronic systems have experienced a rapid development and nowadays are used in a range of applications.[4,5,6] Recent milestones include the use of conventional inorganic magnetic tunnel junctions to fabricate nanoscale nonlinear oscillators[7] and memristive ferroelectric nanodevices[8]. Such breakthroughs bring closer the development of high-density energy-efficient hardware for neuromorphic computing.[9,10,11]

The emergence of molecular spintronics has supposed the extension of the knowledge developed in spintronics with the singular possibilities offered by molecular electronics and molecular magnetism.[12,13,14] These include the ability to design and prepare molecule-based materials *à la carte* that imitate or even enhance the behavior of traditional inorganic spintronic materials. In the road to reduce setup dimensionality, the limit of this paradigm is single-molecule spintronics, which relies on the use of individual molecules as a main device component. A critical step within this framework is the interaction between the magnetic molecule and metallic electrodes.[15] Measurements at a molecular-scale limit are extremely challenging and examples are still scarce, but the first experiments have furnished unique opportunities for exploring quantum effects.[16,17] These seminal works have mainly been based on the second generation of single-molecule magnets, also known as single-ion magnets (SIMs),[18] providing a novel technique for electrically controlling the nuclear spin of a qubit[19] followed by the implementation of Grover's quantum algorithm,[20] between other remarkable achievements.[21,22,23,24] Analogous experiments have also been theoretically explored.[25,26,27]

Among magnetic molecules, lanthanoid SIMs have been the focus of extensive research in the last two decades reaching effective energy barriers above 1800 K[28,29,30] and magnetic hysteresis up to 80 K.[31] The static properties of these minimalistic magnets primarily depend on the magnetic anisotropy of a single ion, which arises from the combination of spin-orbit coupling and an adequate ligand field,[32] whereas the microscopic understanding of their spin dynamics requires to take into account the key role of vibrations that couple to spin states between other factors.[33] The general –and successful– recipe to rationally create $f$-block SIMs is based on the shape of the 4$f$-shell electron density. This results in designing coordination environments that are axially elongated for oblate $f$-ions (e.g. Dy$^{3+}$) and equatorially expanded for prolate ones (e.g. Er$^{3+}$) to stabilize a high-$M_J$ ground state.[34] Whereas the former route has generated most of the SIMs reported in the literature, the latter has been much more limited in practice, due to the preference of lanthanoids for high coordination numbers. This issue was overcome by Tang and co-workers using three bulky ligands to design an exotic equatorially low-coordinated SIM of general formula Er[N(SiMe$_3$)$_2$]$_3$ (**1**) in 2014.[35] More recently, the group of Yamashita has reported slow relaxation of the magnetization in other two three-coordinated Er$^{3+}$ complexes, namely Er(btmsm)$_3$ (**2**) and Er(dbpc)$_3$ (**3**), where btmsm = bis(trimethylsilyl)methyl and dbpc = 2,6-di-*tert*-butyl-*p*-cresolate.[36] The similar $C_{3v}$ symmetry around the magnetic ion but different "soft" C and "hard" O donor atoms offers the advantage to compare their properties and their potential for applications.

Herein, we investigate the charge transport properties of these three low-coordinated SIMs (**1-3**). First, we revisit their electronic structure taking advantage of the experimental energy levels and ligand field analysis reported by Amberger and co-workers.[37,38] Then, we simulate a single-molecule break-junction setup by placing each molecule between two gold electrodes. After a full optimization of molecular structures via density functional theory (DFT) calculations, changes in the coordination environment are analyzed in order to evaluate how magnetic anisotropy may change in the nanodevice. Then, we employ the non-equilibrium Green's function formalism (NEGF) combined with DFT to determine charge and spin electronic transport at low temperatures under the presence of an external magnetic field. Finally, the relations between molecular structures, magnetic properties and transport behavior are discussed.

## 2. Methods

*Crystal field modelling*: The electronic structures of **1**, **2** and **3** were calculated using the latest version of the CONDON package.[39] This software allows the description of the spectroscopic and magnetic properties in $d$ and $f$ systems with high local symmetry using the full basis of microstates. As an input, we have used the reported parameters for the ligand-field Hamiltonian, i.e. interelectronic repulsion parameters $F^k$ ($k$ = 2, 4, 6), spin-orbit coupling constant $\xi_{SO}$, and the non-negligible crystal field parameters (CFPs) for $C_{3v}$ symmetry: $B_{20}$, $B_{40}$, $B_{43}$, $B_{60}$, $B_{63}$ and $B_{66}$. The available $\sigma$ and $\pi$ absorption and luminescence spectra in **1**[37] and **2**[38] have been fitted allowing the code to make a few iterations from the starting point to improve the goodness of the fit (SQX). In the CONDON package, SQX is defined as follows:

$$SQX = \sqrt{\frac{\sum_{i=0}^{n}\frac{1}{\sigma_i}\left(1 - \frac{E_{theo}}{E_{exp}}\right)^2}{n}}$$

where $E_{theo}$ and $E_{exp}$ are the calculated and measured energy levels, respectively, $\sigma_i$ is a weighting factor, which as standard is 1, and $n$ is the number of energy levels included in the fit.

*First-principle calculations*: The *ab initio* calculations were performed using the SMEAGOL code[40] that interfaces the non-equilibrium Green's function (NEGF) approach to electron transport with the density functional theory (DFT) package SIESTA.[41] In all our simulations the transport junction is constructed by placing each molecule between two Au(111)-oriented surfaces with 7x7 cross section. The atomic coordinates of the molecule are relaxed until the total forces are less than 0.01 eV/Å per atom. A real space grid with and equivalent plane wave cutoff of 200 Ry was used to calculate the various matrix elements. Convergence of the electronic structure of the leads was achieved with a 2x2x128 Monkhorst-Pack k-point mesh. Regarding the scattering region, periodic boundary conditions (transverse plane) and open ones (transport direction) are established, using a 2x2x1 k-point mesh. The exchange-correlation potential is described by the Perdew-Burke-Ernzerhof generalized gradient approximation (GGA)[42] (see more details on SI–Section S2).

## 3. Results and discussion

### 3.1. Electronic structure

The three target compounds investigated in this work show a very similar trigonal pyramidal geometry $C_{3v}$, with the erbium cation located slightly out of the plane formed by the three donor atoms (Fig. 1) and dihedral angles of 28.46° (**1**), 37.66° (**2**) and 27.08° (**3**). The high molecular symmetry reduces the number of non-vanishing CFPs, leading to the following CF Hamiltonian (Wybourne formalism)[43]:

$$\hat{H}_{CF}(C_{3v}) = B_{20}C_0^{(2)} + B_{40}C_0^{(4)} + B_{43}(C_{-3}^{(4)} - C_3^{(4)}) + B_{60}C_0^{(6)}$$
$$+ B_{63}(C_{-3}^{(6)} - C_3^{(6)}) + B_{66}(C_{-6}^{(6)} - C_6^{(6)})$$

where $B_{kq}$ are the CFPs and $C_q^{(k)}$ the tensor operators.

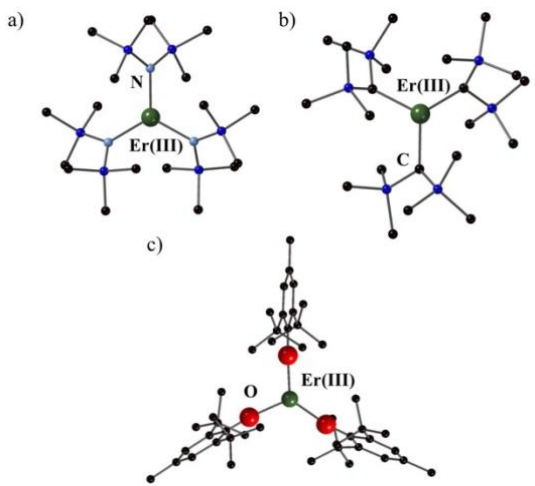

Figure 1: Molecular structures of (a) Er[N(SiMe$_3$)$_2$]$_3$, (b) Er(btmsm)$_3$ and (c) Er(dbpc)$_3$. Hydrogen atoms have been omitted for clarity.

The experimental energy levels of compounds **1** and **2** have been fitted to a full Hamiltonian ($\hat{H} = \hat{H}_{FI} + \hat{H}_{CF}$, where FI denotes free-ion) using the CONDON computational package.[39] In the fitting procedure, only non-vanishing CFPs have been refined in order to improve the phenomenological description. All parameters of the Hamiltonian are reported in Table 1. In both instances, we intended to simultaneously fit the 52 spectroscopic Kramers doublets from different multiplets and the reported magnetic properties,[35,36] but contrary to other works where a number of experimental techniques were available to benchmark and confirm the quality of the data,[44,45] the reported magnetic data in both **1** and **2** was not able to offer a room for improvement with regard to spectroscopic information. To illustrate this, the calculated and experimental $\chi T$ curves are plotted in Figs. S9 and S10. In the case of complex **3**, we were not able to find spectroscopic spectra available in the literature. Nevertheless, the set of non-vanishing CFPs, $F^2$ and $\xi_{SO}$ have been reported for a related compound of general formula Er(OC$_6$H$_3^t$Bu-2,6)$_3$,[38] which only differs from **3** by replacing H by CH$_3$ at the *para-* position of the aromatic ring. This allowed us to provide an estimation of the energy levels, wave functions and magnetic susceptibility of **3**, which could also be compared with the published experimental data (Fig. S11). For this calculation, both $F^4$ and $F^6$ FI terms were extrapolated using the phenomenological values determined for **1** and **2**.

Table 1: Parameters of the Ligand-Field Hamiltonian for **1**, **2** and **3**.

|   | Er[N(SiMe$_3$)$_2$]$_3$ | Er(btmsm)$_3$ | Er(dbpc)$_3$ |
|---|---|---|---|
| $F^2$ | 94772 | 94520 | 95053 |
| $F^4$ | 67699 | 67001 | 68397 |
| $F^6$ | 51052 | 50852 | 51252 |
| $\xi_{SO}$ | 2348 | 2357 | 2347 |
| $B_{20}$ | -2287(122) | -1791(171) | -2095 |
| $B_{40}$ | +341(71) | -66(69) | +735 |
| $B_{60}$ | -101(75) | -53(65) | -49 |
| $B_{43}$ | -725(559) | -7(404) | +45 |
| $B_{63}$ | +52(75) | -89(170) | -581 |
| $B_{66}$ | +196(35) | -504(169) | +118 |
| SQX | 1.90% | 1.86% | – |

In Fig. 2, the calculated ground-*J* multiplet energy levels and their prevailing $M_J$ microstates in the main symmetry ($C_3$) axis are compared for **1**, **2** and **3**. One can observe that the agreement with the experimental energy levels (cross symbols) is excellent. The full set of energy levels is reported as supporting information (Tables ST2-4). In **1** and **2**, the ground state is 99% determined by $|\pm 15/2\rangle$, whereas a mixture of 78% $|\pm 15/2\rangle$ and 21% $|\pm 9/2\rangle$ is observed in **3**. Regarding the separation between the ground and first excited Kramers doublets, we have again a very similar picture. While **1** and **2** have the first excited doublet located at 110 and 117 cm$^{-1}$, respectively, this energy difference is of 78 cm$^{-1}$ in **3**. These descriptions are compatible with the reported single-molecule magnet behavior and agree well with the equatorially-expanded ligand distribution around the metal, but also suggest a slightly unfavored picture –in terms of wave functions mixing and separation between the levels– for **3** with respect to **1** and **2**. Indeed, this trend is to a certain extent reflected in the determined effective energy barriers (U$_{eff}$/k$_B$), which are 122, 114 and 56 K, for **1**, **2** and **3**, respectively. A previous study of these SIMs attributed these differences to the dissimilar donor behavior of C and N with regard to O.[36] The more covalent character of the formers, combined with their similar ligand distribution in the whole molecule, yields a practically analogous distribution of the low-lying energy levels. According to our description, these energy levels are described by wave functions with similar $M_J$ compositions. Other factor that has been discussed by Zhang *et al*.[36] is the average dihedral angles (coordinate vs center-donor planes) based on the experimental chemical structures. This point does not seem to play a major role if we compare the spectroscopic and thermodynamic properties of these molecules, perhaps due to changes in the molecular structure at the range of temperatures in which the experiments are carried out.[46] Of course, a detailed microscopic understanding of the spin dynamics would require to consider spin-phonon interactions,[33] which is beyond the scope of the present work.

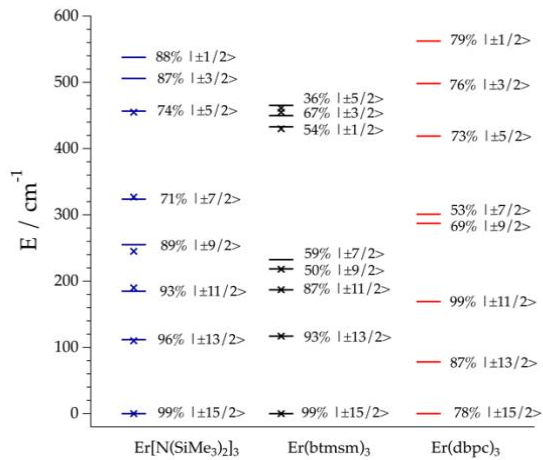

Figure 2: Crystal field energy level scheme of the ground $J$-multiplet of **1** (blue), **2** (black) and **3** (red). Spectroscopic energy levels are marked by a cross.

### 3.2. Structural relaxation in presence of Au electrodes

In a first step to mimic the single-molecule spintronic device, we placed the atomic coordinates of the molecules between two Au (111) electrodes. Then, we evaluated different possibilities for the most relevant parameters of the *in silico* model, which are: (a) *distance between the boundary atoms of the molecules and electrodes*, and (b) *molecular orientation with respect to the Au (111) plane*.

The first point addresses the recovery of an accurate realistic molecular packing in absence of an external strain provoked by the attachment. For that, two possible reasonable molecule-electrode distances were tested, 2.0 and 2.5 Å, obtaining the distance of 2 Å more stable in terms of total energy by more than 2 eV per molecule.

Regarding the conformational orientation of the molecules in the device, two spatial orientations of the triangular plane formed by the donor atoms have been explored. This permits us to test if the substrate could have a major effect on the coordination sphere around the erbium ions. In the first one (*model 1*), the triangular coordination center is aligned parallel to the plane formed by the Au electrodes, while in the second one (*model 2*) the alignment was defined perpendicular to it (Fig. 3). Thus, the magnetic anisotropy easy axis, which coincides with the $C_3$ axis of the molecules, is oriented parallel and perpendicular to the current flow, in *models 1* and *2*, respectively.

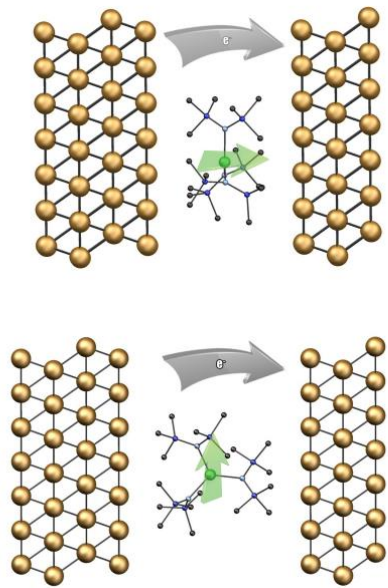

Figure 3: Pictorial representation of the transport single-molecule device. The magnetic anisotropy easy axis is oriented (up) parallel '*model 1*' or (down) perpendicular '*model 2*' with respect to the direction of current flow.

In order to define the most probable thermodynamic structure we have analyzed the relative energies of *models 1* and *2* within the two attachment distances. In all the cases, the parallel configuration is shown to be the most stable in comparison with the perpendicular one. The calculations point a difference between both orientations of 1.582 eV (**1**), 0.376 eV (**2**) and 0.874 eV (**3**). These divergences are not high enough to assure that one model is strongly preferred than the other. Hence, we need to study and compared both scenarios for each molecular device.

The analysis of the geometrical optimization reveals that the structural changes provoked by these structural relaxations are, in all cases, linked to an evolution towards a more symmetric ~$D_{3h}$ coordination environment around the lanthanoid. To illustrate this we can compare, for complex **1**, the azimuthal angle average *($\varphi_{ave}$)*. Whereas in the crystallographic experimental structure $\varphi_{ave}$ = 113.41º, the $\varphi_{ave}$ in the DFT relaxed structures is of 119.99º (*model 1*) and 119.90º (*model 2*). In addition, a reduction in the distance between the lanthanoid and the ligand plane is observed (see Table ST1), which increases the symmetry to a more planar geometry. Considering that for $Er^{3+}$ the quadrupole moment of the *f*-electron charge cloud is prolate (axially elongated), this more equatorially-coordinating geometry will be even more desirable.

*3.3. Spin-transport studies*

We have calculated spin transport properties of each single-molecule spintronic device using **1-3** as a main component. For the two models presented in the previous section, the normalized transmission spectra for spin-up (alpha) and spin-down (beta) has been obtained (Fig. 4). The magnetic moment of the $Er^{3+}$ ion was previously oriented by an external magnetic field. Each transmission spectra shows clear differences for spin-up and spin-down carriers, meaning a preferential transport of one type of carriers than the other. In fact, this was expected due to the presence of a strong magnetic ion with a total spin momentum value J=15/2. This leads to the opportunity of analyzing the relaxation of the magnetic moment by measuring the current passing through the device.

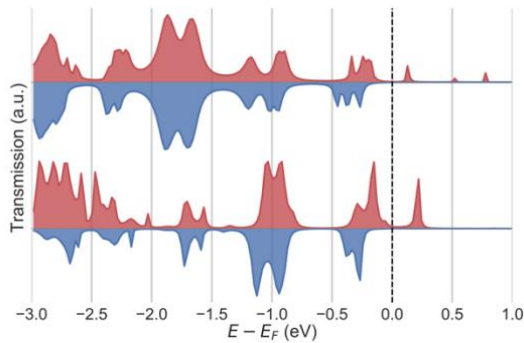

Figure 4: Normalized transmission spectra for **1**, top: parallel conformation (*model 1*), down: perpendicular conformation (*model 2*). Spin-up (red) and spin-down (blue) transmission.

The analysis of transport properties allows us to give insights about how the chemical nature of the ligand is affecting its spin filtering properties. In the case of **1** the transmission spectrum is characterized by a wide distribution of peaks below $E_F$. This is accompanied with a low magnetic influence in the conduction path and the ballistic transport is predicted to be only slightly spin-polarized. The molecular orbitals involved in the transport at $E_F$ are shown in Fig. 5. However, the existence of a spin-up peak at the higher energies (E-$E_F$ ~ 0.15 eV), leads to the possibility of obtaining spin-polarized current by applying a gate voltage to the junction. In this case, *model 2* presents a higher current intensity for the same $V_{gate}$. Finally, the relative absence of transmission peaks at the Fermi level for both *model 1* and *model 2* means that at low bias voltages, the transport through the molecule should occur in a non-polarized tunneling regime.

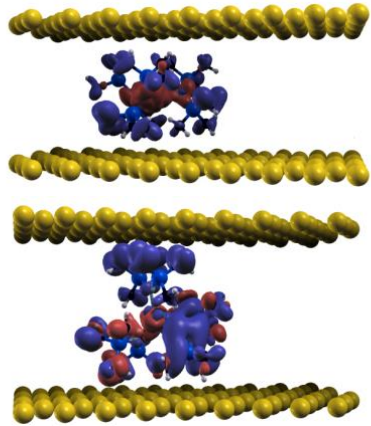

Figure 5: Local density of states at the Fermi level ($E_F$). Up: parallel conformation (*model 1*), down: perpendicular conformation (*model 2*).

For **2**, the calculated transmission spectrum shows several transmission levels with a remarkable spin filtering effect that appear very close to the Fermi level (see SI Fig. S1). In this case, *model 1* presents a spin-up filtering preference while *model 2* allows the spin-down current preferably. In both cases the possibility of obtaining spin-polarized current by applying infinitesimal values of $V_{bias}$ is obtained. Particularly in *model 2*, we also predict an alternation of spin-up and spin-down peaks in the vicinity of $E_F$. As was previously proposed by our group, this opens the door to obtain a double spin filter.[26]

Finally, in **3** we obtain the most significant differences between the parallel and perpendicular configurations. In both cases, the spectrum is characterized by the absence of transmission peaks at $E_F$. Only a few transmission levels between -1.5 and -1 eV in the *model 1* while none is found in *model 2* up to -2.0 eV. (see Fig. S2).

In sum, molecules **1** and **2** which show very similar magnetic core geometry after the relaxation, are expected to have a non-negligible influence of the magnetic ion on molecular filtering properties. In contrast, non-clear spin-dependent peaks were observed along all the spectrum and especially around $E_F$ in **3**, which is geometrically different than the former examples. Considering that C ~ N > O (sorted by ability to form covalent bonds) one may expect a correlation between the electronegativity of the ligand and spin density extension along the molecule, especially in **1** and **2**. We suggest that the covalency of the ligand-metal bond is a key feature while designing spin filter-SIMs.

### 4. Concluding remarks

In summary, we have investigated the transport properties of three equatorially low-coordinated erbium single-ion magnets. The ligand field analysis based on previous spectroscopic energy levels reveals a $M_J = \pm 15/2$ magnetic ground state in **1-3**. By relaxing the molecular structure between two Au (111) electrodes via DFT calculations we obtain a quasi-planar $D_{3h}$ symmetry coordination environment. This supports that single-molecule magnet behaviour is maintained in the proposed nanodevice. In terms of charge-transport, we may expect to use **1** and **2** as basic units for designing molecular spin valves, because of their particular transmission properties. In contrary, a negligible transport value around $E_F$ has been predicted for **3**. This may be assigned to a lower covalent bonding provided by the oxygen atoms. These examples may be used in a new efficient way to study the magnetic relaxation phenomena, as we are able to monitor the spin relaxation by tracking changes in the polarized conductance.

### Acknowledgements

The present work has been funded by the EU (ERC Consolidator Grant 647301 DECRESIM, and COST 15128 Molecular Spintronics Project), the Spanish MINECO (grants MAT2017-89528 and CTQ2017-89993 cofinanced by FEDER and Excellence Unit María de Maeztu MDM-2015-0538), and the Generalitat Valenciana (Prometeo Program of Excellence). S.C.-S. thanks the Spanish MINECO for a Juan de la Cierva-Incorporación postdoctoral Fellowship. J.J.B. thanks the EU for a Marie Curie Fellowship (H2020-MSCA-IF-2016-751047).

1 Baibich, M. N.; Broto, J. M.; Fert, A.; Van Dau, F. N.; Petroff, F.; Etienne, P.; Creuzet, G.; Friederich, A.; Chazelas, J. *Phys. Rev. Lett.* **1988**, *61*, 2472.

2 Binasch, G.; Grünberg, P.; Saurenbach, F.; Zinn, W. *Phys. Rev. B,* **1989**, *39*, 4828.

3 Wolf, S. A.; Awschalom, D. D.; Buhrman, R. A.; Daughton, J. M.; von Molnár, S.; Roukes, M. L.; Chtchelkanova, A. Y.; Treger, D. M. *Science*, **2001**, *294*, 1488-1495.

4 Žutić, I.; Fabian, J.; Das Sarma, S. *Rev. Mod. Phys.*, **2004**, *76*, 323-410.

5 Åkerman, J. *Science*, **2005**, *308*(5721), 508-510.

6 Jungwirth, T.; Sinova, J.; Manchon, A.; Marti, X.; Wunderlich, J.; Felser, C. *Nat. Phys.*, **2018**, *14*, 200-203.

7 Torrejon, J.; Riou, M.; Araujo, F. A.; Tsunegi, S.; Khalsa, G.; Querlioz, D.; Bortolotti, P.; Cros, V.; Yakushiji, K.; Fukushima, A.; Kubota, H.; Yuasa, S.; Stiles, M. D.; Grollier, J. *Nature*, **2017**, *547*, 428-431.

8 Chanthbouala, A.; Garcia, V.; Cherifi, R. O.; Bouzehouane, K.; Fusil, S.; Moya, X.; Xavier, S.; Yamada, H.; Deranlot, C.; Mathur, N. D.; Bibes, M.; Barthélémy, A.; Grollier, J. *Nat. Mater.*, **2012**, *11*, 860–864.

9 Romera, M.; Talatchian, P.; Tsunegi, S.; Araujo, F. A.; Cros, V.; Bortolotti, P.; Trastoy, J.; Yakushiji, K.; Fukushima, A.; Kubota, H.; Yuasa, S.; Ernoult, M.; Vodenicarevic, D.; Hirtzlin, T.; Locatelli, N.; Querlioz, D.; Grollier, J. *Nature*, **2018**, *563*, 230–234.

10 Mizrahi, A.; Hirtzlin, T.; Fukushima, A.; Kubota, H.; Yuasa, S.; Grollier, J.; Querlioz, D. *Nat. Comm.*, **2018**, *9*, 1533.

11 Boyn, S.; Grollier, J.; Lecerf, G.; Xu, B.; Locatelli, N.; Fusil, S.; Girod, S.; Carrétéro, C.; Garcia, K.; Xavier, S.; Tomas, J.; Bellaiche, L.; Bibes, M.; Barthélémy, A.; Saïghi S.; Garcia, V. *Nat. Comm.*, **2017**, *8*, 14736.

12 Rocha, A. R.; García-Suárez, V. M.; Bailey, S. W.; Lambert, C. J.; Ferrer, J.; Sanvito, S. *Nat. Mater.*, **2005**, *4*, 335–339.

13 Bogani, L.; Wernsdorfer, W. *Nat. Mater.*, **2008**, *7*, 179–186.

14 Cornia, A.; Seneor, P. *Nat. Mater.*, **2017**, *15*, 505-506.

15 (a) Godfrin, C.; Thiele, S.; Ferhat, A.; Klyatskaya, S.; Ruben, M.; Wernsdorfer, W.; Balestro, F. *ACS Nano*, **2017**, *11*, 3984. (b) Giménez-Santamarina, S.; Cardona-Serra, S.; Gaita-Ariño, A.; *J. Mag. Mag. Mat.*, **2019**, *485*, 212-216.

.

16 Burzurí, E., van der Zant, H. S. J. Single-Molecule Spintronics. In: Bartolomé J., Luis F., Fernández J. (eds) Molecular Magnets. NanoScience and Technology. Springer, Berlin, Heidelberg (**2014**).

17 Coronado, E.; Yamashita, M.; *Dalton Trans.*, **2016**, *45*, 16553-16555.

18 Ishikawa, N.; Sugita, M.; Ishikawa, T.; Koshihara, S.-y.; Kaizu, Y. *J. Am. Chem. Soc.*, **2003**, *125*, 8694–8695.

19 Thiele, S.; Balestro, F.; Ballou, R.; Klyatskaya, S.; Ruben, M.; Wernsdorfer, W. *Science*, **2014**, *344*, 1135-1138.

20 Godfrin, C.; Ferhat, A.; Ballou, R.; Klyatskaya, S.; Ruben, M.; Wernsdorfer, W.; Balestro, F. *Phys. Rev. Lett.* **2017**, *119*, 187702.

21 Fahrendorf, S.; Atodiresei, N.; Besson, C.; Caciuc, V.; Matthes, F.; Blügel, S.; Kögerler, P.; Bürgler, D. E.; Schneider, C. M. Nat. Commun. 2013, 4, 2425.

22 L. Malavolti, L. Poggini, L. Margheriti, D. Chiappe, P. Graziosi, B. Cortigiani, V. Lanzilotto, F.B. de Mongeot, P. Ohresser, E. Otero, F. Choueikani, P. Sainctavit, I. Bergenti, V.A. Dediu, M. Mannini,R. Sessoli, *Chem. Commun.* **2013**, 49, 11506–11508.

23 Klar, D.; Candini, A.; Joly, L.; Klyatskaya, S.; Krumme, B.; Ohresser, P.; Kappler, J.P.; Ruben, M.; Wende, H. *Dalton Trans.*, **2014**, *43*,10686.

24 Miralles, S. G.; Bedoya-Pinto, A.; Baldoví, J. J.; Cañón-Mancisidor, W.; Prado, Y.; Prima-García, H.; Gaita-Ariño, A.; Mínguez-Espallargas, G.; Hueso, L. E.; Coronado, E. *Chem. Sci.*, **2018**, 9, 199–208.

25 Cardona-Serra, S.; Gaita-Ariño, A; Stamenova, M.; Sanvito, S. *J. Phys. Chem. Lett.*, **2017**, *8*, 3056-3060

26 Cardona-Serra, S.; Gaita-Ariño, A; Navarro-Moratalla, E.; Sanvito, S. *J. Phys. Chem. C*, **2018**, *122*, 6417-6421

27 Cardona-Serra, S.; Gaita-Ariño, A. *Dalton Trans.*, **2018**, *47*, 5533-5537

28 Ding, Y. S.; Chilton, N. F.; Winpenny, R. E. P.; Zheng, Y. Z. *Angew. Chem.* 2016, **55**, 16071–16074

29 Goodwin, C. A.; Ortu, F.; Reta, D.; Chilton, N. F.; Mills, D. P. *Nature* **2017***, 548*, 439

30 Guo, F.-S.; Day, B.; Chen, Y.-C.; Tong, M.-L.; Mansikamäki, A.; Layfield, R. A. *Angew. Chem. Int. Ed.* **2017**, *56*, 11445–11449

31 Guo, F.-S.; Day, B.; Chen, Y.-C.; Tong, M.-L.; Mansikamäki, A.; Layfield, R. A. *Science* **2018**, *362*(6421), 1400-1403.

32 Liddle, S. T.; van Slageren, J. *Chem. Soc. Rev.* **2015***, 44*, 6655–6669

33 (a) Escalera-Moreno, L., Baldoví, J. J., Gaita-Ariño, A.; Coronado, E. *Chem. Sci.*, **2018**, 9, 3265. (b) L. Escalera-Moreno, N. Suaud, A. Gaita-Ariño, E. Coronado, *J. Phys. Chem. Lett.*, 2017, 8, 1695–1700. (c) Lunghi, A.; Totti, F.; Sessoli, R.; Sanvito, S. Nat. Commun. 2017, 8, 14620.

34 (a) Rinehart, J. D.; Long, J. R. *Chem. Sci.* **2011***, 2* (11), 2078. (b) Baldoví, J. J.; Cardona-Serra, S.; Clemente-Juan, J. M.; Coronado, E.; Gaita-Ariño, A.; Palii, A. *Inorg. Chem.* **2012**, *51* (22), 12565–12574.

35 Zhang, P.; Zhang, L.; Wang, C.; Xue, S.; Lin, S.-Y.; Tang, J. *J. Am. Chem. Soc.* **2014**, *136* (12), 4484–4487.

36 Zhang, H.; Nakanishi, R.; Katoh, K.; Breedlove, B. K.; Kitagawa, Y.; Yamashita, M. *Dalton Trans.,* **2018**, 47, 302.

37 Jank, S.; Amberger, H. D.; Edelstein, N. M. *Spectrochim. Acta Part A,* **1998**, *54*, 1645.

38 Reddmann, H.; Guttenberger, C.; Amberger, H.-D. *J. Organomet. Chem.* **2000**, *602*, 65.

39 (a) van Leusen, J.; Speldrich, M.; Schilder, H.; Kögerler, P. *Coord. Chem. Rev.* **2015**, *289*, 137–148. (b) Speldrich, M.; van Leusen, J.; Kögerler, P. *J. Comput. Chem.* **2018**, *39*, 2133.

40 (a) Rocha, A. R.; Garcia-Suarez, V.; Bailey, S. W.; Lambert, C. J.; Ferrer, J.; Sanvito. S. *Nat. Mater.* **2005**, *4*, 335-339. (b) Rungger, I.; Sanvito, S. *Phys. Rev. B,* **2008**, *78*, 035407

41 Soler, M.; Artacho, E.; Gale, J. D.; Garcia, A.; Junquera, J.; Ordejon, P.; Sánchez Portal, D. *J. Phys.: Condens. Matter.* **2002**, *14*, 2745−2779.

42 Perdew, J. P.; Burke, K.; Ernzerhof, M. *Phys. Rev. Lett.* **1996**, *77*, 3865–3868.

43 Wybourne, B. G. Spectroscopic Properties of Rare Earths; John Wiley and Sons, **1965**.

44 Lucaccini, E.; Baldoví, J. J.; Chelazzi, L.; Barra, A. L.; Grepioni, F.; Costes, J. P.; Sorace, L. *Inorg. Chem.*, **2017**, *56*, 4728-4738

45 Coutinho, J.T.; Perfetti, M.; Baldoví, J. J.; Antunes, M. A.; Hallmen, P. P.; Bamberger, H.; Crassee, I.; Orlita, M.; Almeida, M.; van Slageren, J.; Pereira, L. C. J. *Chem.-Eur. J.*, **2019**, *25*(7), 1758-1766.

46 Qian, K.; Baldoví, J. J.; Jiang S.-D.; Gaita-Ariño, A.; Zhang, Y.-Q.; Overgaard, J.; Wang, B.-W.; Coronado, E.; Gao, S. *Chem. Sci.*, **2015**, *6*, 4587-4593

**Supporting Information**

# Exploring the transport properties of equatorially low-coordinated erbium single ion magnets

Silvia Giménez-Santamarina[a], Salvador Cardona-Serra[a*], José J. Baldoví[b*]

[a]*Instituto de Ciencia Molecular, Universitat de València. C/ Catedrático José Beltrán nº 2, 46980 Paterna, Valencia, Spain*

[b]*Max Planck Institute for the Structure and Dynamics of Matter, Luruper Chaussee 149, 22761 Hamburg, Germany*

# S1.- Structural description of the studied SIMs.

Here we report a detailed analysis of the main structural parameters' (angles/distances) of the three SIMs structures after and before the relaxation process for each device geometry.

In all the cases, we can observe an approximation of the metal ion to the ligand plane, this is accompanied with an increase of the average of the angles between the lanthanoid ion and the ligands. This increase pairs with an expansion of the distance between the ligands and the $Er^{3+}$.

Table S1. Structural details of the modeled devices for complexes **1**, **2** and **3**.

|  | $\Theta_1$ (°) | $\Theta_2$ (°) | $\Theta_3$ (°) | $d_{Er\text{-}lig1}$ (Å) | $d_{Er\text{-}lig2}$ (Å) | $d_{Er\text{-}lig3}$ (Å) | $d_{Er\text{-}lig\_plane}$ (Å) |
|---|---|---|---|---|---|---|---|
| **1** – *Initial* | 113.431 | 113.431 | 113.377 | 2.211 | 2.211 | 2.210 | 0.57824 |
| **1** – *model1* | 116.772 | 120.531 | 122.672 | 2.378 | 2.374 | 2.370 | 0.04715 |
| **1** – *model2* | 122.492 | 125.617 | 111.601 | 2.375 | 2.375 | 2.372 | 0.12374 |
| **2** – *Initial* | 108.163 | 108.164 | 108.164 | 2.358 | 2.358 | 2.358 | 0.83529 |
| **2** – *model1* | 122.113 | 117.677 | 120.189 | 2.590 | 2.581 | 2.595 | 0.02084 |
| **2** – *model2* | 124.265 | 122.970 | 111.278 | 2.561 | 2.571 | 2.570 | 0.20809 |
| **3** – *Initial* | 112.683 | 115.247 | 114.290 | 2.040 | 2.048 | 2.041 | 0.50644 |
| **3** – *model1* | 111.517 | 128.038 | 120.406 | 2.173 | 2.181 | 2.168 | 0.10323 |
| **3** – *model2* | 123.990 | 126.061 | 109.925 | 2.182 | 2.175 | 2.179 | 0.10870 |

# S2.- Extended theoretical methodology

Transport calculations are performed using the SMEAGOL code[1] which includes non-equilibrium Green's function (NEGF) approach to the density functional theory (DFT) package SIESTA[2]. Structural optimization calculation was performed using the original SIESTA code.

In all our simulations the transport junction is constructed by placing the molecule between two Au(111)-oriented surfaces with 7x7 cross section. This mimics a standard transport break-junction experiment with the most used gold surface orientation.

Thus, the structures are relaxed until the maximum forces are less than 0.01 eV/Å. In the optimization, dispersion corrections were not included because the expected weak interaction between the coordinating ligands and the electrodes. A real space grid with and equivalent plane wave cutoff of 200 Ry (enough to ensure convergence) has been used to calculate the various matrix elements. Finally, the electronic temperature of the calculation (unless specified) is set to 0.1 K to mimic the low-temperature limit conditions used in the original experiment. During the calculation, the total system is divided in three parts: a left-hand side lead, a central scattering region (SR) and a right-hand side lead. The scattering region contains the molecule as well as four atomic layers of each lead, which are necessary to relax the electrostatic potential to the bulk level of Au. Due to the difficulties to obtain an adequate pseudopotential (appropriate relationship between accuracy and time consumption) in these big ensembles we performed the relaxation calculations substituting the magnetic $Er^{3+}$ ion by a $Ca^{2+}$ ion. We ensure the correspondence by a charge adjustment and considering comparable coordination properties due to the similarity on the ionic radii.

The convergence of the electronic structure of the leads is achieved with 2x2x128 Monkhorst-Pack k-point mesh, while for the SR one sets open boundary conditions in the transport direction and periodic ones along the transverse plane, for which an identical k-point mesh is used (2x2x1 k-points). The exchange-correlation potential is described by the Perdew-Burke-Ernzerhof (PBE) generalized gradient approximation (GGA).[3] Owing to the quasi-pure $M_J = \pm 15/2$ wave function of the ground state in our systems, with negligible contributions from other $M_J$ states, the mono-determinantal DFT approach can be considered as valid approximation in this case.

The Au-valence electrons are represented over a numerical s-only single-θ basis set that has been previously demonstrated to offer a good description of the energy region around the Fermi level[4]. In contrast, for the other atoms we use a full-valence double-θ basis set with polarization (basis size was increased until convergence). Norm-conserving Troullier-Martins pseudopotentials[5] are employed to describe the core-electrons in all the cases.

Finally, the spin-dependent current, $I_\sigma$, flowing through the junction is calculated from the Landauer-Büttiker formula[6],

$$I_\sigma(V) = \frac{e}{h} \int_{-\infty}^{+\infty} T_\sigma(E,V)[f_L(E - \mu_L) - f_R(E - \mu_R)]dE$$

Where the total current $I_{tot}$, is the sum of both the spin-polarized components, $I_\sigma$, where $\sigma$ = spin-up/spin-down. Here $T_\sigma(E,V)$ is the transmission coefficient[1], and $f_{L/R}$ are the Fermi functions associated with the two electrodes' chemical potentials, $\mu_{L/R} = \mu_\sigma \pm V/2$, where $\mu_\sigma$ is the electrodes' common Fermi level.

In our two-spin-fluid approximation (there is no spin-flip mechanism) majority and minority spins carry two separate spin currents, and the resultant current spin polarization, SP, is calculated as

$$SP = \frac{I_{up} - I_{down}}{I_{up} + I_{down}}$$

## S3.- Calculated transmission spectra for the complexes 2 and 3.

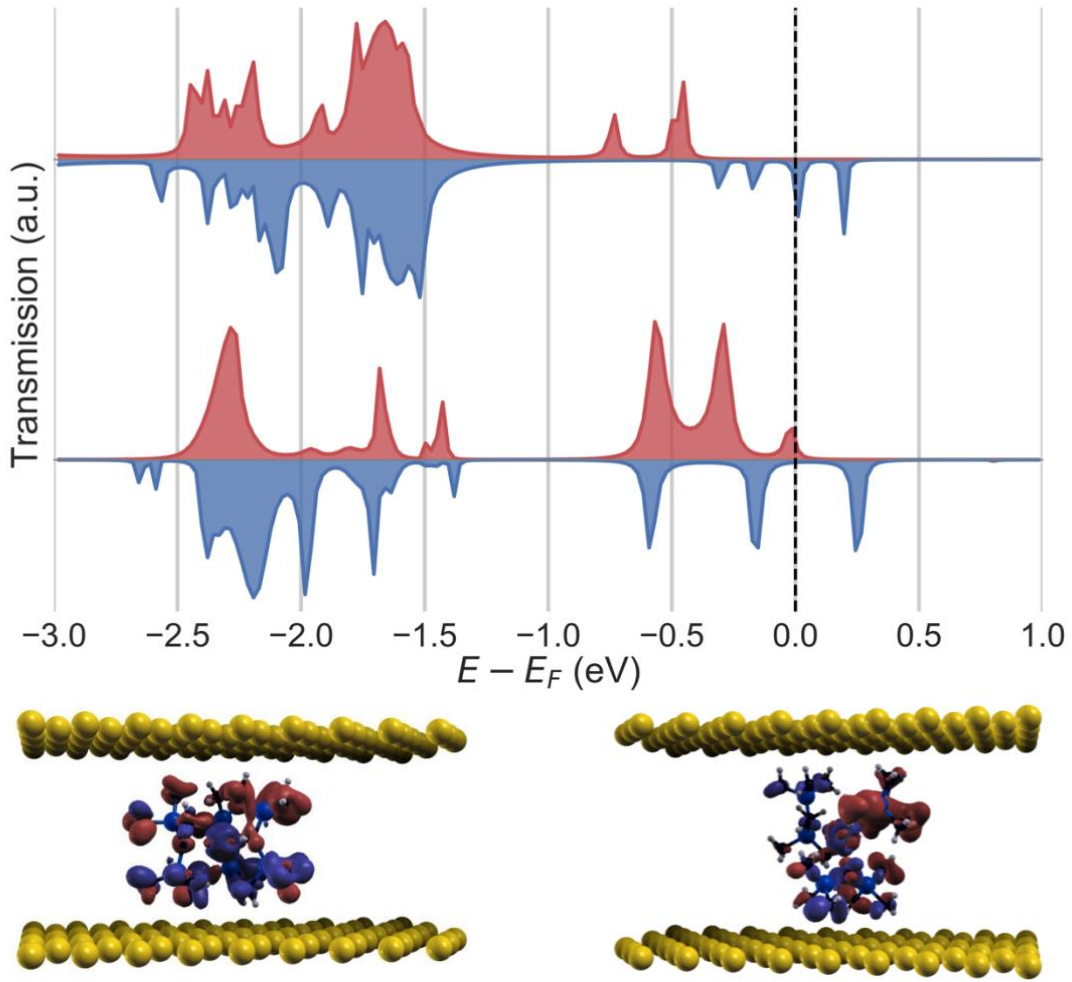

Figure S1: Upper panel: Normalized transmission spectra for complex **2**, top: parallel conformation (*model 1*), down: perpendicular conformation (*model 2*). (Red: spin-up transmission, Blue: spin-down transmission). Lower panel: Local density of states at the Fermi level ($E_F$) (left: parallel conformation, right, perpendicular conformation).

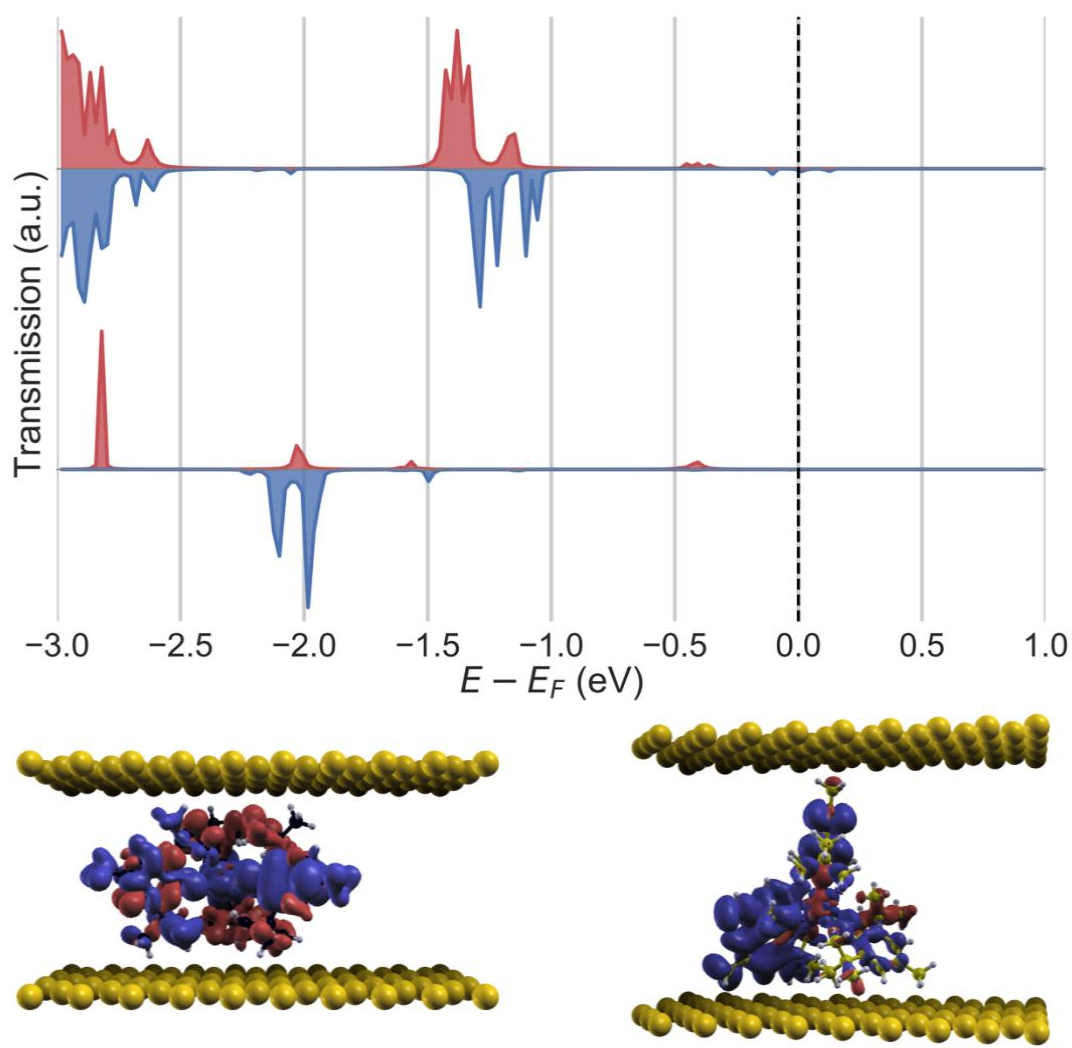

Figure S2: Upper panel: Normalized transmission spectra for complex **3**, top: parallel conformation (*model 1*), down: perpendicular conformation (*model 2*). (Red: spin-up transmission, Blue: spin-down transmission). Lower panel: Local density of states at the Fermi level ($E_F$) (left: parallel conformation, right, perpendicular conformation).

# S4.- Relaxed structures of the complexes in the single-molecule transport device.

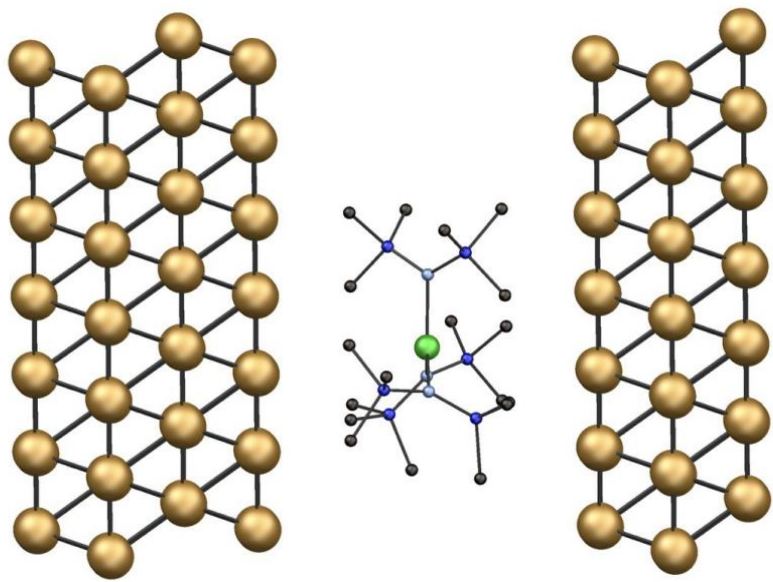

Figure S3: Ball&Stick representation of the scattering region for *model 1* in complex **1**, H atoms have been removed for clarity.

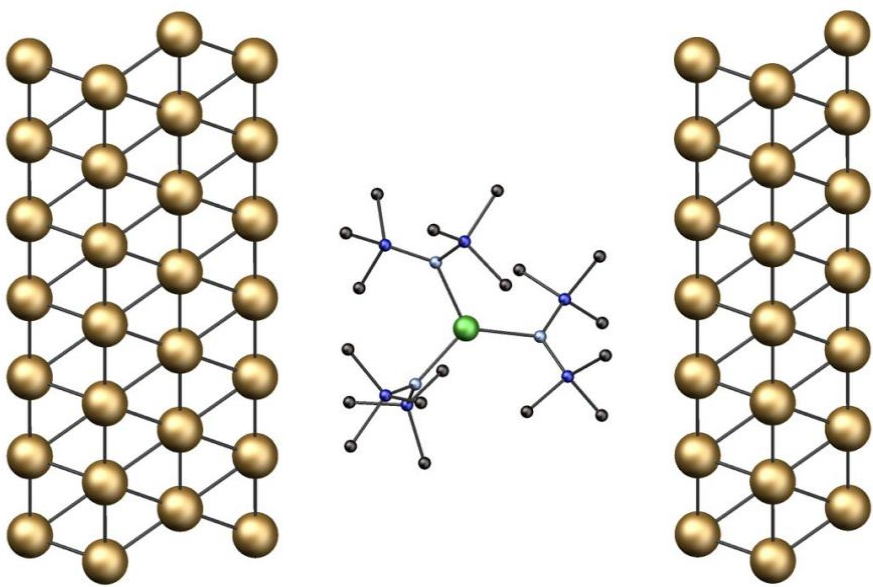

Figure S4: Ball&Stick representation of the scattering region for *model 2* in complex **1**, H atoms have been removed for clarity.

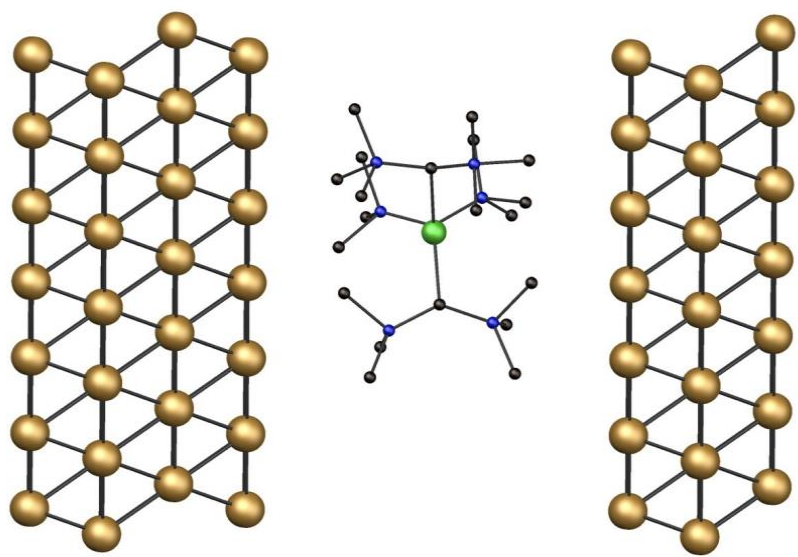

Figure S5: Ball&Stick representation of the scattering region for *model 1* in complex **2**, H atoms have been removed for clarity.

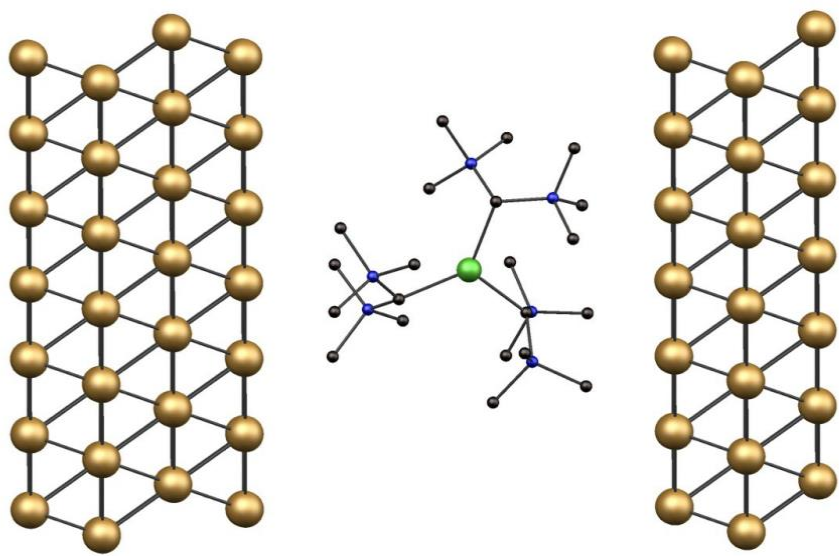

Figure S6: Ball&Stick representation of the scattering region for *model 2* in complex **2**, H atoms have been removed for clarity.

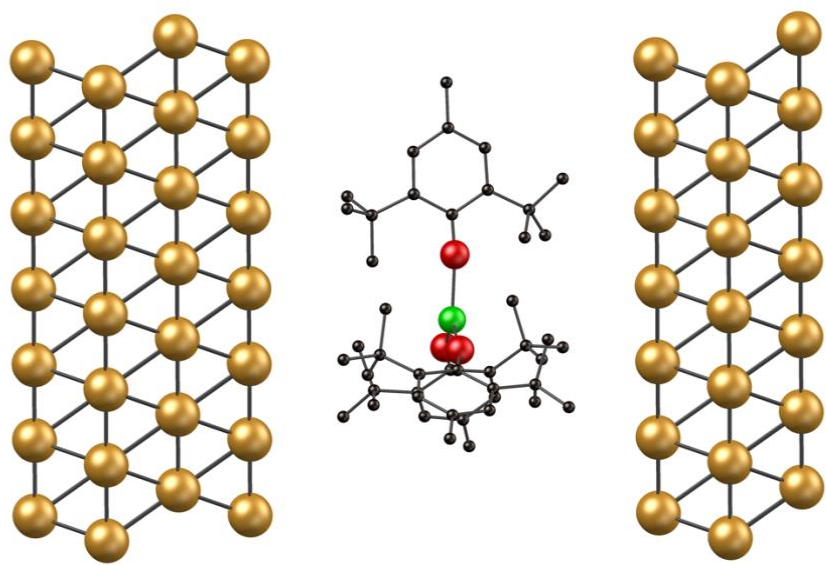

Figure S7: Ball&Stick representation of the scattering region for *model 1* in complex **3**, H atoms have been removed for clarity.

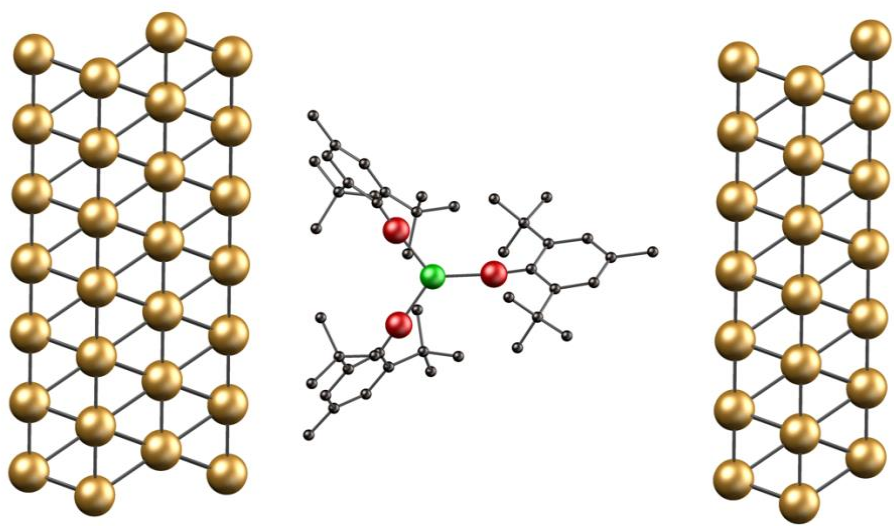

Figure S8: Ball&Stick representation of the scattering region for *model 2* in complex **3**, H atoms have been removed for clarity.

## S5.- Magnetic susceptibility simulations using the full Hamiltonian

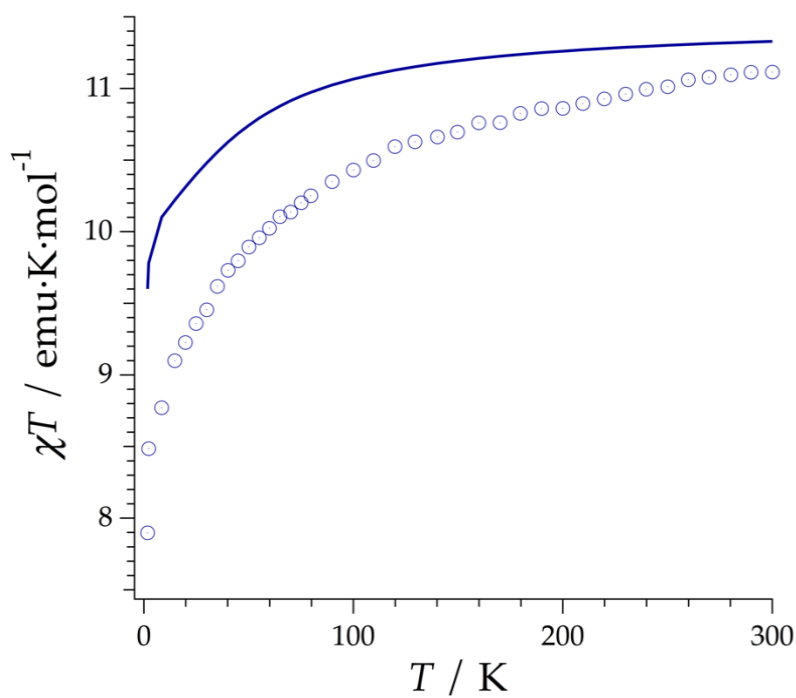

Figure S9: Experimental (symbols), calculated (solid line) temperature-dependence of the powder magnetic susceptibility of complex **1** from 2 to 300 K measured at 1 kOe.

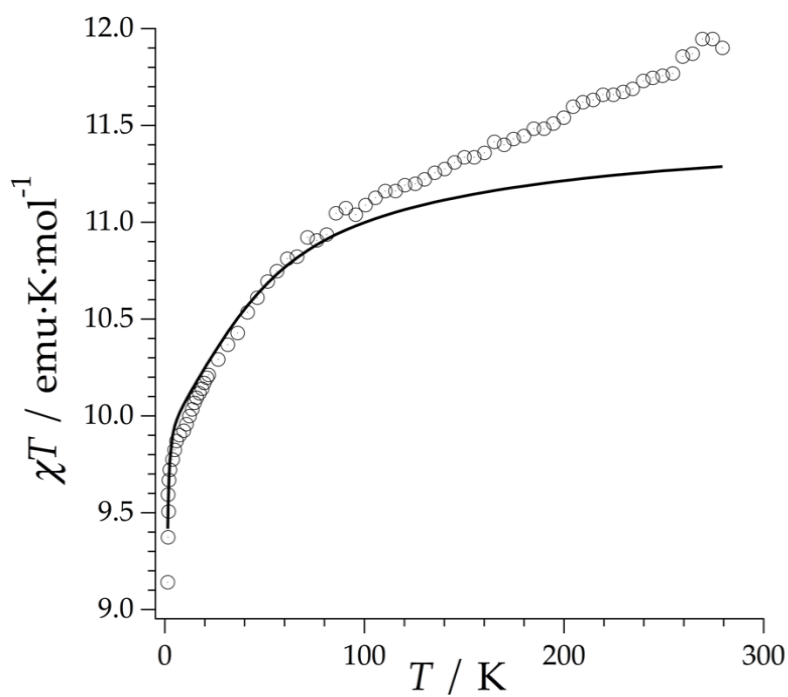

Figure S10: Experimental (symbols), calculated (solid line) temperature-dependence of the powder magnetic susceptibility of a diluted analogue of complex **2**, $Er_{0.05}Y_{0.95}(btmsm)_3$, from 2 to 300 K measured at 1 kOe.

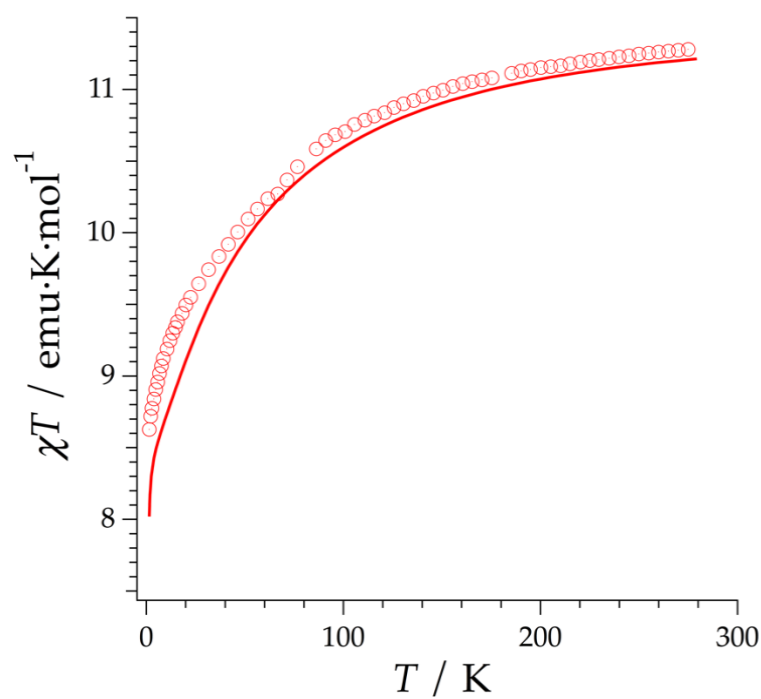

Figure S11: Experimental (symbols), calculated (solid line) temperature-dependence of the powder magnetic susceptibility of complex **3** from 2 to 300 K measured at 1 kOe.

[1] (a) Rocha, A. R.; Garcia-Suarez, V.; Bailey, S. W.; Lambert, C. J.; Ferrer, J.; Sanvito. S. Towards Molecular Spintronics. *Nat. Mater.* **2005**, *4*, 335-339. (b) Rungger, I.; Sanvito, S. Algorithm for the Construction of Self-Energies for Electronic Transport Calculations Based on Singularity Elimination and Singular Value Decomposition. *Phys. Rev. B* **2008**, *78*, 035407

[2] Soler, M.; Artacho, E.; Gale, J. D.; Garcia, A.; Junquera, J.; Ordejon, P.; Sánchez Portal, D. The SIESTA Method for Ab-Initio Order-N Materials Simulation. *J. Phys.: Condens. Matter* **2002**, *14*, 2745−2779.

[3] Perdew, J. P.; Burke, K.; Ernzerhof, M. *Phys. Rev. Lett*. **1996**, *77*, 3865–3868.

[4] Toher, C.; Sanvito, S. Effects of Self-Interaction Corrections on the Transport Properties of Phenyl-Based Molecular Junctions. *Phys. Rev. B*, **2008**, *77*, 155402.

[5] Troullier, N.; Martins, J. L. Efficient Pseudopotentials for Plane-Wave Calculations. *Phys. Rev. B* **1991**, *43*, 1993-2006.

[6] Büttiker, M.; Imry, Y.; Landauer, R.; Pinhas, S. Generalized Many-Channel Conductance Formula with Applications to Small Rings. *Phys. Rev. B: Condens. Matter Mater. Phys.* **1985**, *31*, 6207−6215.